\documentclass[superscriptaddress,nofootinbib,notitlepage,aps,prd]{revtex4-2}
\usepackage{amsfonts}
\usepackage[bookmarksopen,colorlinks]{hyperref}
\usepackage{graphicx}
\usepackage{dcolumn}
\usepackage{bm}
\usepackage{amssymb,amsmath}
\usepackage{mathtools}
\usepackage[dvips]{color}
\usepackage[utf8]{inputenc}
\usepackage[title]{appendix}

\begin{document}

\title{Noether's second theorem in teleparallel gravity}
\author{Rafael Ferraro}

\altaffiliation[]{ferraro@iafe.uba.ar} \affiliation{Instituto de
Astronom\'\i a y F\'\i sica del Espacio (IAFE, CONICET-UBA), Casilla
de Correo 67, Sucursal 28, 1428 Buenos Aires, Argentina.}
\affiliation{Departamento de F\'\i sica, Facultad de Ciencias
Exactas y Naturales, Universidad de Buenos Aires, Ciudad
Universitaria, Pabell\'on I, 1428 Buenos Aires, Argentina.}

\begin{abstract}
Gauge symmetries in teleparallel gravity, together with the
identities among the dynamical equations they provide, are analyzed
in relation to the way they condition the coupling between matter
and gravity. Particularly, the coupling of fermionic matter seems to
be excluded in a wide range of teleparallel theories.
\end{abstract}

\maketitle

\section{Introduction}

One of the distinctive features of Maxwell and Einstein equations is the
\textit{automatic conservation},
\begin{equation}
\partial_\mu\partial_\nu(\sqrt{|g|}~F^{\mu \nu })\equiv 0\ ,
~~~~~~~~~~G_{\ ~;\;\mu}^{\mu \nu }\equiv 0\ ,  \label{automatic00}
\end{equation}
which are off-shell identities evidencing that the dynamical
equations are not independent; they imply a restriction on the
number of genuine degrees of freedom (dof) the dynamical equations
govern. Noether's second theorem traces the automatic conservation
to a gauge symmetry of the action. The electromagnetic Lagrangian
$L_{em}\propto |g|^{1/2} F^{\mu\nu}F_{\mu\nu}$ is made of the field
$F_{\mu\nu}=\partial_\mu A_\nu-\partial_\nu A_\mu$ which is
invariant under gauge transformations
\begin{equation}
A_{\mu }\rightarrow A_{\mu }+\partial_\mu \xi \ . \label{gauge}
\end{equation}
Thus the action is insensitive to the \textit{local} variations
$\delta A_{\mu }=\partial_\mu\xi$, not only on the solution to the
dynamical equations but on arbitrary field evolutions. These
variations are unable to generate dynamics; as a consequence, the
resulting dynamical equations cannot be independent. Function $\xi
(x)$ in Eq.~(\ref{gauge}) is called the \textit{generator} of the
gauge transformation. The gauge freedom of the potential $A_{\mu }$
can be completely fixed by means of two gauge conditions (for
instance, $\vec{\nabla}\cdot \vec{A}=0$ and $A_{0}=0$). Then the
gauge symmetry (\ref{gauge}) involves two spurious dof, so leaving
the potential with two genuine dof (transversal modes). The reason
why two spurious dof are associated with just one gauge generator
comes from the fact that $\partial _{0}A_{0}$ is not present in
$F_{\mu\nu}$. Thus $A_{0}$ lacks a kinetic term in the Lagrangian,
which means that $A_{0}$ is not a genuine dof. Besides, since
$A_{0}$ is differentiated only along spatial directions, the
variation of the action with respect to $A_{0}$ does not result in a
dynamical equation for other components of $A_\mu$ but in the
\textit{constraint} $\vec{\nabla}\cdot \vec{E}=\rho /\epsilon _{o}$,
which implies one additional spurious dof. So, let us review how the
automatic conservation $(\sqrt{|g|}~F^{\mu \nu })_{,\nu\mu}\equiv 0$
emerges from the gauge symmetry. The general variation of an action
$S[A_\mu]$ is
\begin{equation}
\delta S=\int d^{4}x\ \left[ \frac{\delta L}{\delta A_{\mu }}
~\delta A_{\mu }+\partial _{\nu }\left( \frac{\partial L}{\partial
(\partial _{\nu }A_{\mu })}\ \delta A_{\mu }\right) \right]
\label{variationEM}
\end{equation}
where $\delta L/\delta A_{\mu }$ stands for the Euler-Lagrange
derivative $\partial L/\partial A_{\mu }-\partial _{\nu }\left(
\partial L/\partial (\partial _{\nu }A_{\mu })\right) $. For the
gauge symmetry (\ref{gauge}), it is $\delta A_{\mu }=\partial_\mu
\xi$ and $\delta S_{em}\equiv 0$; thus, after an integration by
parts one gets\footnote{To write the rhs~of Eq.~(\ref{00}), we have
used that $\partial L_{em}/\partial (\partial _{\nu }A_{\mu })$ is
antisymmetric in $\mu ,\nu $.}
\begin{equation}
\int d^{4}x\ \xi \ \partial _{\mu }\left( \frac{\delta
L_{em}}{\delta A_{\mu }}\right) \equiv \int d^{4}x\ \partial _{\mu
}\left[ \xi \
\partial _{\nu }\left( \frac{\partial L_{em}}{\partial (\partial
_{\nu }A_{\mu })}\right) +\xi \ \frac{\delta L_{em}}{\delta A_{\mu
}}\right] ~.  \label{00}
\end{equation}
If the symmetry (\ref{gauge}) were valid only for a specific
function $\xi (x)$, then Eq.~(\ref{00}) would express a conservation
law on-shell (i.e., when the Euler-Lagrange equations $\delta
L_{em}/\delta A_{\mu }=0$ are fulfilled). The bracket in the rhs~of
Eq.~(\ref{00}) would be then the (divergenceless) current associated
with the conserved charge, as stated in Noether's first theorem.
However the gauge symmetry (\ref{gauge}) is valid no matter what the
function $\xi (x)$ is. This essential feature puts the
Eq.~(\ref{00}) in the context of Noether's second theorem. While the
lhs~of Eq.~(\ref{00}) depends on the behavior of an arbitrary
function $\xi (x)$ in the entire region of integration, the rhs~just
depends on its behavior on the boundary. Therefore, Eq.~(\ref{00})
makes sense only if the integrands on each side are identically zero
\textit{off-shell}. The conservation law we could deduce in this
context is, to use Emmy Noether's words, \textit{improper
}\cite{Noether}: the zero divergence of the current in the rhs~of
Eq.~(\ref{00}) is not a consequence of the dynamics but a mere
off-shell identity.\footnote{See Noether's analysis of Hilbert's
assertion about the non-existence of a proper concept of
gravitational energy in general relativity. Noether points out that
the space-time translations in general relativity are just a
subgroup of an infinity dimensional symmetry group (diffeomorphisms)
for which her second theorem applies.} The off-shell vanishing of
$\partial_\mu(\delta L_{em}/\delta A_{\mu })$ in Eq.~(\ref{00})
yields the automatic conservation of Eq.~(\ref{automatic00}). On the
other hand, if the field is not free but sources are present, then
the automatic conservation will force them to be conserved.

\bigskip

In the next Sections we will discuss gauge symmetries in theories of
gravity that admit a teleparallel formulation, the identities at the
level of the dynamical equations they provide, and the way in which
these identities restrict the coupling of the matter to gravity.
Sections \ref{II} and \ref{III} are devoted to the invariance
associated with diffeomorphisms. Section \ref{IV} considers the
invariance of general relativity under local Lorentz transformations
of the tetrad. Section \ref{V} studies the compatibility of the
Dirac field with the identities coming from the gauge symmetries.
Section \ref{VI} displays the conclusions.

\section{Teleparallel gravity}\label{II}

General relativity (GR) also exhibits a gauge symmetry; the
transformation of the metric tensor \cite{Sundermeyer}
\begin{equation}
\mathbf{g}\rightarrow \mathbf{g}+\pounds _{\mathbf{\xi }}~\mathbf{g}\ ,~~~~\
~~~~~\text{or}~~~\ ~~~~~g_{\mu \nu }\rightarrow g_{\mu \nu }+2\;\xi _{(\mu
;\nu )}\ ,  \label{EHgauge}
\end{equation}
changes the Einstein-Hilbert action by a boundary term; so these
variations of the metric do not generate dynamics. Thus, Einstein
equations are not independent but are linked by the automatic
conservation (\ref{automatic00}). Since the gauge transformations
(\ref{EHgauge}) come with four arbitrary generator functions $\xi
_{\mu }(x)$, they suppress eight spurious dof; thus only two genuine
dof~remain among the ten components of the metric.\footnote{In
Eq.~(\ref{EHgauge}) $\pounds$ denotes the Lie-derivative; the
semicolon stands for the Levi-Civita covariant derivative. In a
chart where the Levi-Civita connection locally vanishes, it is
$g_{\mu \nu }\rightarrow g_{\mu \nu }+2\; g_{\lambda (\mu } \xi_{\;,
\nu )}^{\lambda }$, which coincides with the infinitesimal
transformation of the metric components under the change of
coordinates $x^{\lambda }\rightarrow x^{\lambda}+\xi^{\lambda
}$.\label{notediff}}

General relativity and other theories of gravity can be formulated
in terms of the tetrad field $\{\mathbf{E}^{a}(x)\}$ --the basis of
the cotangent space-- and its dual basis $\{\mathbf{e}_{a}(x)\}$,
\begin{equation}
e_{a}^{\mu }~E_{\nu }^{a}=\delta _{\nu }^{\mu }~,~~~~e_{a}^{\mu }~E_{\mu
}^{b}=\delta _{a}^{b}\ .  \label{dual}
\end{equation}%
The tetrad relates to the metric through the orthonormality condition
\begin{equation}
g_{\mu \nu }~e_{a}^{\mu }~e_{b}^{\nu }=\eta _{ab}~,~~~~g_{\mu \nu }=\eta
_{ab}~E_{\mu }^{a}~E_{\nu }^{b}\ ,  \label{metric}
\end{equation}%
where $\eta _{ab}=\mathit{diag}\{1,-1,-1,-1\}$ is the Minkowski
symbol.

Like $\mathbf{F}=d\mathbf{A}$ in electromagnetism,
$\mathbf{T}^{a}=d\mathbf{E}^{a}$ is the fundamental magnitude in
\textit{teleparallel gravity}.  Although $\mathbf{T}^{a}$ is
invariant under the transformations $\mathbf{E}^a\rightarrow
\mathbf{E}^a+d\xi^a$, this is not a symmetry in teleparallel
theories because teleparallel Lagrangians are also made of vectors
$\{\mathbf{e}_{a}\}$ and the volume $E=\det [E_{\mu
}^{a}]=|g|^{1/2}$. The set of four 2-forms $\mathbf{T}^{a}$ can be
read as a \textit{torsion}, provided the Weitzenb\"{o}ck connection
$\Gamma _{\lambda\mu }^{\rho }=e_{b}^{\rho }~E_{\mu ,\lambda }^{b}$
is adopted:
\begin{equation}
T_{\ \lambda \mu }^{\rho }=e_{a}^{\rho }\,T_{\ \lambda\mu
}^{a}=e_{a}^{\rho }~(\partial_\lambda E_{\mu}^{a}-\partial_\mu
E_{\lambda}^{a})=\Gamma _{\lambda \mu }^{\rho}-\Gamma _{\mu \lambda
}^{\rho}~.
\end{equation}%
Weitzenb\"{o}ck connection proves to be flat (so, parallelism is \textit{%
absolute}; it does not depend on the path), and cancels out the
covariant derivative of the tetrad (then, it is a \textit{metric}
connection).

\bigskip

The \textit{teleparallel equivalent of general relativity} (TEGR) is
defined by Lagrangian density
\begin{equation}
L_{TEGR}=(2\kappa)^{-1}~E~S_{\rho }^{~\lambda \mu }~T_{~\lambda \mu
}^{\rho }\ ,  \label{TEGR}
\end{equation}
where $S_{\rho }^{~\lambda\mu}$ is
\begin{equation}
2\,S_{\rho }^{~\lambda \mu }=\underbrace{\frac{1}{2}\,(T_{\rho }^{\
\lambda \mu }-2~T_{\ \ \ \ \rho }^{[\lambda \mu
]}}_{\text{\textit{contorsion}}\ K_{\ \ \,\rho }^{\lambda \mu
}})+2~T^{[\lambda }\,\delta _{\rho }^{\mu ]}\ , \label{S}
\end{equation}%
and $T^{\mu }=T_{\lambda }^{\ \,\lambda \mu }=K_{\ \ \,\lambda
}^{\lambda \mu }$ is the \textit{torsion vector}. On the other hand,
the Einstein-Hilbert Lagrangian is $L_{EH}=-(2\kappa)^{-1}
|g|^{1/2}R$, where the Levi-Civita curvature $R$ depends on second
derivatives of the metric. TEGR and GR are dynamically equivalent
because, once the relation (\ref{metric}) is used to write $R$ in
terms of the tetrad, their Lagrangians differ in a divergence
\begin{equation}
L_{TEGR}=L_{EH}+\partial _{\nu }(\kappa ^{-1}E~T^{\nu })\ .
\label{HE}
\end{equation}

Any teleparallel theory like TEGR, whose Lagrangian density $L$ is
homogeneous of degree $2$ in first derivatives of the tetrad,
accepts the following form for $L$:
\begin{equation}
L=\frac{1}{4 \kappa}~E\ E_{\,\,\nu ,\mu }^{a}\ E_{\,\,\lambda ,\rho
}^{b}\ e_{c}^{\mu }~e_{e}^{\nu }~e_{d}^{\rho }~e_{f}^{\lambda
}\,M_{ab}^{\,\,\,\,\,cedf}\,.  \label{LagrM}
\end{equation}%
In TEGR the symbol $M_{ab}^{\,\,\,\,cedf}$ is \cite{Ferraro:2016}
\begin{equation}
M_{ab}^{\,\,\,\,cedf}=2\,\eta _{ab}\,\eta ^{c[d}\,\eta
^{f]e}-4\,\delta _{a}^{[d}\,\eta ^{f][c}\,\delta _{b}^{e]}+8\,\delta
_{a}^{[c}\,\eta ^{e][d}\,\delta _{b}^{f]}\,,  \label{supermetric}
\end{equation}%
but other combinations of its three terms are proposed in
alternative theories of gravity known as New General Relativity
(NGR)
\cite{Guzman:2020egp,Hayashi,Cheng:1988zg,Blixt:2018znp,Blixt:2020ekl}.
Note that the antisymmetrized indices of $M_{ab}^{\,\,\,\,cedf}$
make the derivatives of the tetrad enter the Lagrangian only through
the components of the Weitzenb\"{o}ck torsion.

Lagrangian density (\ref{LagrM}) takes its most elegant form when
written in terms of the anholonomy or commutation coefficients
$f_{bc}^{a}$,
\begin{equation}
\pounds _{\mathbf{e}_{b}}\mathbf{e}_{c}\equiv \lbrack
\mathbf{e}_{b},\mathbf{e}_{c}]=f_{bc}^{a}~\mathbf{e}_{a}
~~~~~~~\Rightarrow~~~~~~~f_{bc}^{a}=-(\pounds
_{\mathbf{e}_{b}}\mathbf{E}^{a})(\mathbf{e} _{c})=-2~e_{c}^{\mu
}e_{b}^{\nu }~E_{[\mu ,\nu ]}^{a}=e_{c}^{\mu }e_{b}^{\nu
}~T^a_{\;\mu\nu} \label{fabc}
\end{equation}
(Eq.~(\ref{dual}) is used to solve $f_{bc}^{a}$). Thus the
Lagrangian is written in terms of scalar objects valued in the
tangent space,
\begin{equation}
L=\frac{1}{16 \kappa} ~E\
f_{ce}^{a}~f_{df}^{b}~M_{ab}^{\,\,\,\,\,cedf}:=(2 \kappa)^{-1} E~T\
, \label{LagrM2}
\end{equation}%
since no coordinate indices are left in its constituent parts (cf. Ref.~%
\cite{Cho,Aldrovandi:2013wha}).

Teleparallel dynamics is invariant under the infinitesimal local
transformation (the analogous of (\ref{EHgauge}))
\begin{equation}
\mathbf{E}^{a}\rightarrow \mathbf{E}^{a}+\pounds _{\mathbf{\xi }}\mathbf{E}%
^{a}\ ,~~~~~~~~~~~~~~\mathbf{e}_{a}\rightarrow \mathbf{e}_{a}+\pounds _{\mathbf{\xi }}\mathbf{e}%
_{a}=\mathbf{e}_{a}+[\mathbf{\xi },\mathbf{e}_{a}]\ ,
\label{tetradgauge}
\end{equation}%
which keeps the duality (\ref{dual}). In fact, on the one hand
Eq.~(\ref{fabc}) implies that \footnote{Lie and exterior derivatives
commute.}
\begin{equation}
\delta f_{bc}^{a}=\delta\left(e_{c}^{\mu }e_{b}^{\nu
}~(d\mathbf{E}^a)_{\mu\nu}\right)=\pounds _{\mathbf{\xi
}}f_{bc}^{a}=\xi ^{\mu } f_{bc ,\mu }^{a}\ ;
\end{equation}
then it is $\delta T=\xi ^{\mu }T_{,\mu }$. On the other hand, the
change of the volume is
\begin{equation}
\delta E=E~e_{a}^{\lambda }~\delta E_{\lambda }^{a}=E~e_{a}^{\lambda
}~(\xi^\nu E_{\lambda ,\nu}^{a}+E_{\nu }^{a}~\xi^\nu_{,\lambda})=(E
\xi^\nu)_{,\nu}\ .\label{deltaE}
\end{equation}
The behavior of the Lagrangian (\ref{LagrM2}) under the
transformation (\ref{tetradgauge}) is then
\begin{equation}
\delta L=\delta (E\ T)=(E~\xi ^{\nu }T)_{,\nu }=(L~\xi ^{\nu })_{,
\nu }~. \label{deltaLagr}
\end{equation}%
Equation (\ref{deltaLagr}) implies the action gets a boundary term
under the transformation (\ref{tetradgauge}), so the dynamics is not
affected. This conclusion can be extended to other teleparallel
formulations, like $f(T)$ gravity; in fact,
\begin{equation}
\delta \lbrack E~f(T)]~=f(T)~\delta E+E~f^{\prime }(T)~\delta
T=f(T)~(E \xi ^{\nu })_{,\nu }+E~f^{\prime }(T)~\delta T=(E~f(T)~\xi
^{\nu })_{,\nu }~.
\end{equation}%
The result $\delta L=(L\;\xi^{\nu})_{,\nu }$ is common to any theory
of the spacetime geometry whose Lagrangian scalar is made of a
chart-independent combination of the commutation coefficients. Thus
teleparallel actions are not strictly gauge invariant under the
transformation (\ref{tetradgauge}), but $\delta S$ turns out to be
\footnote{$\delta E_{\mu }^{a}$ in the Euler-Lagrange derivative
involves the differentiation of $e_{b}^{\nu }$. Duality (\ref{dual})
implies that $e_{a}^{\nu }\;\delta E_{\mu }^{a}=-E_{\mu
}^{a}\;\delta e_{a}^{\nu }$.\label{v}}
\begin{equation}
\delta S\vert_{\delta E_{\mu }^{a}=(\pounds _{\mathbf{\xi
}}\mathbf{E}^{a})_{\mu }}=\int d^{4}x\ \left[ \frac{\delta L}{\delta
E_{\mu }^{a}}~\delta E_{\mu }^{a}+\partial _{\nu }\left(
\frac{\partial L}{\partial (\partial _{\nu }E_{\mu }^{a})}\ \delta
E_{\mu }^{a}\right) \right] _{\delta E_{\mu }^{a}=(\pounds
_{\mathbf{\xi }}\mathbf{E}^{a})_{\mu }}=\int d^{4}x~\partial _{\nu
}(L~\xi ^{\nu })~,
\end{equation}
i.e.,
\begin{equation}
\int d^{4}x\ \frac{\delta L}{\delta E_{\mu }^{a}}~(\pounds
_{\mathbf{\xi }}\mathbf{E}^{a})_{\mu }=\int
d^{4}x~\mathrm{divergence~.}\label{interm}
\end{equation}
The components of $\pounds _{\mathbf{\xi }}\mathbf{E}^{a}$ can be
written in different ways,
\begin{equation}
(\pounds _{\mathbf{\xi }}\mathbf{E}^{a})_{\mu }=\xi ^{\lambda
}~E_{\mu ,\lambda }^{a}+E_{\lambda }^{a}~\xi _{~,\mu }^{\lambda
}=(E_{\lambda }^{a} \xi ^{\lambda })_{,\mu }+\xi ^{\lambda }(E_{\mu
,\lambda }^{a}-E_{\lambda ,\mu }^{a})=(E_{\lambda }^{a} \xi
^{\lambda })_{,\mu }+\xi^{\lambda} T_{\,\,\lambda \mu }^{a}\ ;
\label{tetradgauge2}
\end{equation}%
we will replace the last one in Eq.~(\ref{interm}) which, after
integration by parts, turns out to be
\begin{equation}
\int d^{4}x\ \xi ^{\lambda }\left( -E_{\lambda }^{a}\;\partial _{\mu
}+T_{~\lambda \mu }^{a}\right) \,\frac{\delta L}{\delta E_{\mu
}^{a}}=\int d^{4}x~\mathrm{divergence~.}  \label{variationTele}
\end{equation}
In this equation the lhs depends on the behavior of the
\textit{arbitrary} infinitesimal vector field $\mathbf{\xi }$ in the
entire region of integration, while the rhs only depends on the
behavior of $\mathbf{\xi }$ and its first derivatives at the
boundary. Therefore, Eq.~(\ref{variationTele}) makes sense only if
both integrands are identically zero off-shell, which leads to the
automatic conservation
\begin{equation}
\left( E_{\lambda }^{a}\;\partial _{\mu }-T_{~\lambda \mu
}^{a}\right) ~ \frac{\delta L}{\delta E_{\mu }^{a}} \equiv 0\ .
\label{automatic}
\end{equation}%
By rearranging this result, or using the first of the forms of $(\pounds _{\mathbf{%
\xi }}\mathbf{E}^{a})_{\mu }$ in Eq.~(\ref{tetradgauge2}), we can also write
\begin{equation}
\partial _{\mu }\left( E_{\lambda }^{a}\;\frac{\delta L}{\delta E_{\mu }^{a}}
\right)-E_{\mu ,\lambda }^{a}~\frac{\delta L}{\delta E_{\mu
}^{a}}\equiv 0 \ .  \label{automatic2}
\end{equation}
These identities about the tensor density $E_{\lambda }^{a}~\delta
L/\delta E_{\mu }^{a}$, which possesses the structure of the
dynamical equations, express the content of Noether's second theorem
in teleparallel gravity regarding the gauge symmetry
(\ref{tetradgauge}). They can be used to write the Levi-Civita
divergence of the tensor $E^{-1}E_{\lambda }^{a}~\delta L/\delta
E_{\mu }^{a}$, to give the identity a more familiar look. We will
need the relation between Weitzenb\"{o}ck and Levi-Civita connections,
$\Gamma $ and $\big\{\ \big\}$, which is given by the contorsion
tensor introduced in Eq.~(\ref{S}):
\begin{equation}
K_{~\mu \lambda }^{\rho }=\Gamma _{\lambda \mu }^{\rho
}-\genfrac{\{}{\}}{0pt}{0}{\rho}{\lambda\mu}~~~~~~~~\Rightarrow
~~~~~~~~\genfrac{\{}{\}}{0pt}{0}{\rho}{\lambda\mu}=e_b^\rho~
E^b_{\mu,\lambda}-K_{~\mu \lambda }^{\rho}\
,~~~~\genfrac{\{}{\}}{0pt}{0}{\mu}{\lambda\mu}=e_b^\mu~
E^b_{\mu,\lambda}=E^{-1}\partial_\lambda E~, \label{contorsion}
\end{equation}
since $K_{~\mu \lambda }^{\mu}=0$. Thus, the Levi-Civita divergence
of a tensor is
\begin{equation}
C_{~\lambda ;\mu }^{\mu }=\partial _{\mu }C_{~\lambda }^{\mu
}+\genfrac{\{}{\}}{0pt}{0}{\mu}{\mu\rho}~C_{~\lambda }^{\rho
}-\genfrac{\{}{\}}{0pt}{0}{\rho}{\mu\lambda}~C_{~\rho }^{\mu
}=E^{-1}~\partial _{\mu }(E~C_{~\lambda }^{\mu })-(e_b^\rho~
E^b_{\mu,\lambda}-K_{~\mu \lambda }^{\rho })~C_{~\rho }^{\mu }~,
\label{WDiv0}
\end{equation}
where we have used the symmetry of the Levi-Civita connection. By
replacing the tensor density $E C_{~\lambda }^{\mu }$ with
$E_{\lambda }^{a}~\delta L/\delta E_{\mu }^{a}$, and using
Eq.~(\ref{automatic2}) to cancel out terms, one obtains
\begin{equation}
\left( E^{-1}E_{\lambda }^{a}\;\frac{\delta L}{\delta E_{\mu
}^{a}}\right) _{;\;\mu }- K_{~\mu \lambda }^{\rho }~E^{-1}~E_{\rho
}^{a}\;\frac{\delta L}{\delta E_{\mu }^{a}}\equiv 0~,
\label{automatic22}
\end{equation}
which is an equivalent way of writing the automatic conservation
(\ref{automatic2}).

\section{The action of matter}\label{III}

The automatic conservation is not only important as a tracer of the
spurious dof~in a theory. It also forces the sources to obey a
conservation law with its same form. In electrodynamics, the
automatic conservation $(\sqrt{|g|}~F^{\mu \nu
})_{,\nu \mu }\equiv 0$ constrains the charges to be conserved: $(%
\sqrt{|g|}~j^{\mu })_{,\mu }=0$. In turn, charge conservation is the
key to have the gauge invariance of the coupling term
$\sqrt{|g|}~A_{\mu }~j^{\mu }$ in the full field-charge action,
since it changes as
\begin{equation}
\sqrt{|g|}~A_{\mu }~j^{\mu }\rightarrow \sqrt{|g|}~(A_{\mu }+\xi
_{,\mu
})~j^{\mu }=\sqrt{|g|}~A_{\mu }~j^{\mu }-\xi ~(\sqrt{|g|}~j^{\mu })_{,\mu }+%
\mathrm{divergence}~.
\end{equation}%
So, the conservation of the sources and the gauge symmetry of the coupling
term are two features that go together.

In GR, the automatic conservation $G_{~~;\;\mu }^{\mu \nu }\equiv 0$
forces the energy-momentum tensor of the sources to satisfy
$\mathfrak{T}_{~~;\mu }^{\mu \nu }=0$. This means that the automatic
conservation in theories of gravity essentially determines the
dynamics of the sources; it fixes the evolution of the matter
energy-momentum in each geometry.\footnote{See \S 20.6 in
Ref.~\cite{Gravitation} for nice examples and discussions about this
issue.} Like in electromagnetism, the coupling matter-gravity must
exhibit the symmetry that gives rise to the automatic conservation;
only then will the dynamics of the sources be compatible with the
dictates of the automatic conservation. This is a severe restriction
to the form of the action of matter, which is entirely a
matter-gravity coupling action because it is necessarily formulated
in a geometric background. This symmetry requirement is overcome by
writing the action of matter in terms of geometric objects, as is
the case with the gravity action. In other words the action must be
invariant under diffeomorphisms (see Note \ref{notediff}). Actually
the action of matter must come with \textit{all} the gauge
symmetries of the gravity action to provide a consistent set of
dynamical equations.

In the case of the teleparallel gauge symmetry (\ref{tetradgauge}),
the action of matter must satisfy Eq.~(\ref{automatic22}) on-shell.
It is expected that the matter will keep to this consistency
requirement by conserving its energy-momentum. This is so for any
type matter that couples to the metric but not to its derivatives,
like scalar or spin 1 matter. In fact, the absence of derivatives of
the tetrad implies that
\begin{equation}
E_{\lambda }^{a}\;\frac{\delta L_{mat}}{\delta E_{\mu
}^{a}}=E_{\lambda }^{a}\;\frac{\partial L_{mat}}{\partial E_{\mu
}^{a}}=E_{\lambda }^{a}\;\frac{\partial L_{mat}}{\partial g_{\rho
\nu }}\ \frac{\partial g_{\rho \nu }}{\partial E_{\mu
}^{a}}=-\sqrt{|g|}\ \mathfrak{T}^{\rho \nu }\,\eta _{ab}\,~\delta
_{\rho }^{\mu }~E_{\lambda }^{a}\;E_{\nu }^{b}=-E\
\mathfrak{T}_{~\lambda }^{\mu }\ .  \label{matConserv}
\end{equation}
where $\mathfrak{T}^{\mu \lambda }= -2~|g|^{-1/2}~\partial
L_{mat}/\partial g_{\mu \lambda }$ is the \textit{metric}
energy-momentum tensor (the source of Einstein equations). By
replacing this result in Eq.~(\ref{automatic22}) one gets
\begin{equation}
\mathfrak{T}_{~\lambda ;\mu }^{\mu }-K_{~\mu \lambda }^{\rho
}~\mathfrak{T}_{~\rho }^{\mu }=0~.  \label{automatictmunu}
\end{equation}%
At first sight this result seems to contradict the equivalence
principle, since one expects to recover the Special-Relativity law
$\mathfrak{T}_{~\lambda ,\mu }^{\mu }=0$ in a chart where the
Levi-Civita connection locally vanishes and the freely falling
particles locally move on straight curves (their parametric
equations are linear in the affine parameter). However the issue is
solved by noting that the contorsion tensor is antisymmetric in its
two first indices ($K_{~~~\lambda }^{\mu \nu }=-K_{~~~\lambda }^{\nu
\mu }$) and the metric energy-momentum tensor is symmetric. Thus the
second term on the lhs of Eq.~(\ref{automatictmunu}) is zero, and
the equivalence principle is safe (cf. Ref.~\cite{Hayashi}).

\section{Lorentz gauge invariance in TEGR}\label{IV}

The antisymmetric part of TEGR dynamical equations is identically
zero, as stressed in several publications
\cite{Hayashi,Golovnev:2017dox,Golovnev:2018gk,Ong,Golovnev:2020nln}.
This is an identity exclusive to TEGR, that comes from the TEGR
gauge invariance under local Lorentz transformation of the tetrad;
so it also falls within the framework of Noether's second theorem.
As an object of the vector representation
$\left(\frac{1}{2},\frac{1}{2}\right)$ of the Lorentz group, the
tetrad in Eq.~(\ref{metric}) transforms as
\begin{equation}
\mathbf{E}^{a^{\prime }}=\Lambda _{~a}^{a^{\prime }}~\mathbf{E}^{a}~,~~~\
~~~~~~~\mathbf{e}_{a}=\mathbf{e}_{a^{\prime }}~\Lambda _{~a}^{a^{\prime
}}~,~~~\ ~~~~~~~\Lambda _{~a}^{a^{\prime }}=\exp \left[ \frac{1}{2}~\sigma
_{gh}~(M^{gh})_{~a}^{a^{\prime }}\right] ~,  \label{VecTransf}
\end{equation}
where the six generators $M^{gh}$ are
\begin{equation}
(M^{gh})_{~b}^{a}=2~\eta ^{a[g}\ \delta _{b}^{h]}~.  \label{VecGen}
\end{equation}%
If the parameters $\sigma _{gh}=\sigma _{\lbrack gh]}$ are
infinitesimal, the Lorentz transformation becomes
\begin{equation}
\delta _{L}E_{\mu }^{a}=\sigma _{gh}~\eta ^{a[g}\ \delta _{b}^{h]}~E_{\mu
}^{b}=\sigma _{~b}^{a}~E_{\mu }^{b}~,~~\ \ ~\ \ ~~\ \ \delta _{L}e_{b}^{\nu
}=-e_{a}^{\nu }~\sigma _{gh}~\eta ^{a[g}\ \delta _{b}^{h]}=-e_{a}^{\nu
}~\sigma _{~b}^{a}~.  \label{deltaL}
\end{equation}%
Einstein-Hilbert action is invariant under \textit{local} Lorentz
transformations of the tetrad (parameters $\sigma _{gh}$ become
arbitrary functions) because it depends purely on the components of
the metric (\ref{metric}), which are locally Lorentz invariant.
Instead TEGR action is \textit{pseudo-invariant}, since $S_{TEGR}$
gets a boundary term after the transformation due to the divergence
term in Eq.~(\ref{HE}). The pseudo-invariance implies that the
general structure of $\delta_L S_{TEGR}$ (see
Eqs.~(\ref{variationEM}) and (\ref{variationTele})) is not
identically zero in this case, but it is
\begin{equation}
\delta _{L}S_{TEGR}=\int d^{4}x\left[ \frac{\delta L_{TEGR}}{\delta
E_{\mu }^{a}}\;\delta _{L}E_{\mu }^{a}+\partial _{\nu }\left(
\frac{\partial L_{TEGR}}{\partial (\partial _{\nu }E_{\mu
}^{a})}~\delta _{L}E_{\mu }^{a}\right) \right] \equiv \delta
_{L}\int d^{4}x~\partial _{\nu }(\kappa ^{-1}ET^{\nu })~.
\label{variations}
\end{equation}%
Therefore it results
\begin{equation}
\int d^{4}x~\frac{\delta L_{TEGR}}{\delta E_{\mu }^{a}}~\delta
_{L}E_{\mu }^{a}~\equiv \int d^{4}x~\partial _{\nu }\left( \kappa
^{-1}E~\delta
_{L}T^{\nu }-\frac{\partial L_{TEGR}}{\partial (\partial _{\nu }E_{\mu }^{a})}%
~\delta _{L}E_{\mu }^{a}\right)  \label{variationLorentz}
\end{equation}%
(Lorentz matrices are unimodular, so $E=\det E_{\mu }^{a}$ is
Lorentz invariant). The lhs~in Eq.~(\ref{variationLorentz}) depends
on the behavior of the \textit{free} parameters $\sigma_{gh}(x)$ in
the entire region of integration; instead, the rhs~depends on their
values at the boundary. This result makes sense only if each side
identically vanishes whatever the parameters $\sigma_{gh}$ are. Thus
we obtain six off-shell identities among the TEGR dynamical
equations
\begin{equation}
\frac{\delta L_{TEGR}}{\delta E_{\mu }^{a}}~\eta ^{a[g}~E_{\mu
}^{h]}\equiv 0\ . \label{TEGRas}
\end{equation}%
Equation (\ref{TEGRas}) shows that the antisymmetric part of TEGR
dynamical equations is identically zero. The way the matter couples
to gravity in TEGR (and GR) must be consistent with this identity.
The full action $S_{TEGR}+S_{matter}$ must be (pseudo-) invariant
under local Lorentz transformations of the tetrad field, otherwise
the dynamical equations would be sourced by an (incompatible)
non-symmetric energy-momentum tensor. As shown above, this point is
not a problem for matter fields that couple to the metric.

Unlike gauge transformations (\ref{gauge}) and (\ref{tetradgauge}),
the transformation (\ref{deltaL}) does not contain derivatives of
the parameters $\sigma ^{ag}$. Therefore, the identities
(\ref{TEGRas}) do not have the appearance of automatic conservation;
they are six relations among the TEGR dynamical equations (not their
derivatives). Neither the counting of the spurious dof develops like
in the electromagnetic case, since the TEGR Lorentz gauge symmetry
does not come from the absence of a kinetic term. The six off-shell
identities (\ref{TEGRas}) actually suppress six spurious dof, which
must be added to the eight spurious dof coming from the automatic
conservation (like in electromagnetism, each component of
$\mathbf{\xi }$ in Eq.~(\ref{tetradgauge}) involves two spurious
dof). In sum, the sixteen elements of the tetrad $E_{\mu }^{a}$
contain only $16-2\times 4-6=2$ genuine dof.

\section{Dirac field}\label{V}

We will now consider the Dirac field, which couples not to the
metric but to the tetrad field. So, we will study how the Dirac
field accommodates to the identities coming from the gravity sector.
Dirac Lagrangian
\begin{equation}
L_{D}=E~(i~\overline{\psi }~\gamma ^{c}e_{c}^{\nu }~\partial _{\nu
}\psi -m \overline{\psi }\psi )  \label{LD}
\end{equation}
is a scalar density under changes of chart, since $\psi $ behaves
like a scalar field in such case, and is invariant under
\textit{global} Lorentz transformations. As an object of the $\left(
\frac{1}{2},0\right) \oplus \left( 0,\frac{1}{2}\right) $
representation of the Lorentz group, the Dirac spinor has four
complex components $\psi ^{\alpha }$ which transform as \cite{Tong}
\begin{equation}
\psi ^{\alpha ^{\prime }}=U_{~\alpha }^{\alpha ^{\prime }}~\psi ^{\alpha
},~~~\ ~~~~\overline{\psi }_{\beta ^{\prime }}=(\psi ^{\dagger }\gamma
^{0})_{\beta ^{\prime }}=~\overline{\psi }_{\beta }~U_{~\beta ^{\prime
}}^{\beta }~,\ ~~~~\mathrm{where}~~~~\ ~~~U_{~\alpha ^{\prime }}^{\beta
}~U_{~\alpha }^{\alpha ^{\prime }}=\delta _{\alpha }^{\beta }~,~~~\
~~~~U_{~\alpha }^{\alpha ^{\prime }}=\exp \left[ \frac{1}{2}~\sigma
_{gh}~(S^{gh})_{~\alpha }^{\alpha ^{\prime }}\right] ~.~
\end{equation}%
$\sigma _{gh}=\sigma _{\lbrack gh]}$ are the six parameters
characterizing the set of Lorentz transformations.\footnote{$\alpha
$, $\beta $ are spinor labels. They are also labels for the
components of the Dirac matrices $\gamma ^{c}$, which have been
omitted.} The generators $S^{gh}$ are
\begin{equation}
S^{gh}=\frac{1}{4}~[\gamma ^{g},\gamma ^{h}]~,  \label{SpinorGen}
\end{equation}%
and Dirac matrices $\gamma ^{a}$ satisfy the Clifford algebra%
\begin{equation}
\{\gamma ^{a},\gamma ^{b}\}=2~\eta ^{ab}~\mathbf{1}~. \label{Diracm}
\end{equation}%
As generators of different representations of the same group,
$S^{gh}$ in Eq.~(\ref{SpinorGen})~and $M^{gh}$ in Eq.~(\ref{VecGen})
both obey the Lorentz algebra. The same parameters $\sigma _{gh}$
must be used to have the same Lorentz transformation in each
representation. Thus, those parameters used to rotate a vector
$360^o$ make the spinor to change the sign. The invariance of Dirac
equation
\begin{equation}
i~\gamma ^{c}e_{c}^{\nu }~\partial _{\nu }\psi -m\psi =0~,~~~\ ~~~\
\ ~~~~\ ~i~e_{c}^{\nu }~\partial _{\nu }\overline{\psi }~\gamma
^{c}+m\overline{\psi }=0~,\label{Diraceq}
\end{equation}
under Lorentz transformations implies that the Dirac matrices
transform as~\cite{Gasperini}
\begin{equation}
\gamma ^{c^{\prime }}=U~\gamma ^{c}~\Lambda _{~c}^{c^{\prime
}}~U^{-1}~,~~~\ \ \ \ ~~~~\mathrm{or}~~~\ \ \ \ ~~~~\Lambda
_{~c^{\prime }}^{d}~\gamma ^{c^{\prime }}=U~\gamma ^{d}~U^{-1}~,
\label{gammaTransf}
\end{equation}
which moves the Dirac matrices to another representation of the
Clifford algebra (\ref{Diracm}). This means that the change of basis
$\Lambda$ in the tangent space is compensated by the respective
linear transformation $U$ in the space of spinors, so leaving the
Dirac equation invariant.

Let us compute $E_{\lambda }^{a}\;\delta L_{D}/\delta E_{\mu }^{a}$
to test the consistency of Dirac action with the automatic
conservation (\ref{automatic2}). Since $L_D$ does not contain
derivatives of the tetrad, it is \footnote{We have used $\partial
E/\partial E_{\mu }^{a}=E~e_{a}^{\mu }$, and $\partial e_{b}^{\nu
}/\partial E_{\mu }^{a}=-e_{b}^{\mu }~e_{a}^{\nu }$ (see Note
\ref{v}).}
\begin{equation}
E_{\lambda }^{a}\;\frac{\delta L_{D}}{\delta E_{\mu
}^{a}}~=E_{\lambda }^{a}\;\frac{\partial L_{D}}{\partial E_{\mu
}^{a}}~=~L_{D}~\delta _{\lambda }^{\mu}-E~i~\overline{\psi }~\gamma
^{c}~e_{c}^{\mu }~\partial_{\lambda }\psi=-E~\mathfrak{T}_{~\lambda
}^{\mu } ~,\label{DiracAut2}
\end{equation}
where $\mathfrak{T}_{~\lambda }^{\mu }$ is the canonical
energy-momentum tensor of the Dirac field. The automatic
conservation (\ref{automatic2}) compels the Dirac dynamics to make
zero the quantity $(E~\mathfrak{T}_{~\lambda }^{\mu
})_{,\mu}-E^a_{\mu,\lambda} ~e_a^\rho~ E~\mathfrak{T}_{~\rho }^{\mu
}$. By using Eq.~(\ref{Diraceq}) it becomes
\begin{equation}
(E~\mathfrak{T}_{~\lambda }^{\mu })_{,\mu}-E^a_{\mu,\lambda}
~e_a^\rho~ E~\mathfrak{T}_{~\rho }^{\mu }=i~\overline{\psi }~\gamma
^{c} \left(E~e_{c}^{\mu }\right)_{,\mu} \partial_{\lambda
}\psi~.\label{result}
\end{equation}
Therefore, the compatibility with the automatic conservation could
be reached in a gauge fixed scheme where the Lorenz gauge
$\left(E~e_{c}^{\mu }\right)_{,\mu}=0$ (partially) fixes the gauge
freedom (\ref{tetradgauge}).\footnote{$(E  e_{c}^{\mu})_{,\mu} = E ~
e_{c\; ;\mu}^{\mu}$ relates to the torsion vector since
$T_\lambda=-E^c_\lambda ~ e_{c\; ;\mu}^{\mu}$.}

\bigskip

Tetrad $\mathbf{e}_c=e_{c}^{\nu} \partial_{\nu}$ is the geometric
object (independent of the chart) that provides the Dirac equation
(\ref{Diraceq}) with the information about the
gravitational-inertial field, thus taking the role that the metric
plays for bosonic fields. If gravity is governed by TEGR (or GR)
dynamics, then the tetrad will be determined \textit{modulo} local
Lorentz transformations. However Dirac Lagrangian (\ref{LD}) is not
invariant under local Lorentz transformations because $\partial
_{\nu }\psi$ does not transform as a spinor. Therefore local Lorentz
transformations are not allowed in Dirac theory unless a covariant
derivative $D _{\nu }\psi$ be introduced. The covariant derivative
$D _{\nu }\psi$ must transform as a spinor,
\begin{equation}
(D_{\nu }\psi )^{\alpha ^{\prime }}=U_{~\alpha }^{\alpha ^{\prime
}}~(D_{\nu }\psi )^{\alpha }~,\label{cdtransf}
\end{equation}
which will require a (to be determined) connection term
\cite{Fock},\footnote{For a historical account see \cite{Kay}. For
fermion coupling in the broader context of general (linear) affine
geometries see \cite{Beltran}.}
\begin{equation}
D_{\nu }\psi =\left( \partial _{\nu }+\frac{1}{2}~\Omega _{gh\nu
}~S^{gh}\right) \psi\ .\label{cd}
\end{equation}
Thus,
\begin{equation}
(D_{\nu }\psi )^{\prime }=\left( \partial _{\nu }+\frac{1}{2}~\Omega
_{g^{\prime }h^{\prime }\nu }~S^{g^{\prime }h^{\prime }}\right) \psi
^{\prime }=U~\left( \partial _{\nu }+U^{-1}(\partial _{\nu
}U)+\frac{1}{2}~\Omega _{g^{\prime }h^{\prime }\nu
}~U^{-1}S^{g^{\prime }h^{\prime }}U\right) ~\psi ~.
\end{equation}%
According to Eq.~(\ref{gammaTransf}) it is
\begin{equation}
S^{g^{\prime }h^{\prime }}=\Lambda _{~g}^{g^{\prime }}~\Lambda
_{~h}^{h^{\prime }}~U~S^{gh}~U^{-1}~,
\end{equation}
therefore
\begin{equation}
(D_{\nu }\psi )^{\prime }=U~\left( \partial _{\nu }+U^{-1}(\partial
_{\nu }U)+\frac{1}{2}~\Omega _{g^{\prime }h^{\prime }\nu }~\Lambda
_{~g}^{g^{\prime }}~\Lambda _{~h}^{h^{\prime }}~S^{gh}\right) ~\psi
~.  \label{CovDer}
\end{equation}
To satisfy Eq.~(\ref{cdtransf}) the parenthesis in the rhs~should be
equal to $D_{\nu }$. For infinitesimal transformations we have
\begin{equation}
U^{-1}(\partial _{\nu }U)=\frac{1}{2}~\sigma _{gh,\nu }~S^{gh}~.
\end{equation}
This shows that a proper connection in Eq.~(\ref{CovDer}) is one
that transforms as
\begin{equation}
\Omega _{g^{\prime }h^{\prime }\nu }~\Lambda _{~g}^{g^{\prime
}}~\Lambda _{~h}^{h^{\prime }}=\Omega _{gh\nu }-\sigma _{gh,\nu
}~,~~~~\ ~~\mathrm{i.e.,}~~\ \ ~~~~~~~\delta _{L}\Omega _{gh\nu
}=-\sigma _{gh,\nu }~. \label{connection00}
\end{equation}
We will try with
\begin{equation}
\Omega _{gh\nu }=-\eta _{d[g}~e_{h]}^{\rho }~E_{\rho ,\nu }^{d}+~...
\label{connection0}
\end{equation}%
since, according to Eq.~(\ref{VecTransf}), it fulfills the
Eq.~(\ref{connection00}):
\begin{equation}
~\delta _{L}(-\eta _{d[g}~e_{h]}^{\rho }~E_{\rho ,\nu }^{d})=-\eta
_{d[g}~e_{h]}^{\rho }~(\delta _{L}E_{\rho }^{d})_{,\nu }=-\eta
_{d[g}~e_{h]}^{\rho }~\sigma _{~b,\nu }^{d}~E_{\rho }^{b}=-\sigma
_{gh,\nu }~.
\end{equation}%
Equation (\ref{connection0}) only shows the compensation term we
need to build a covariant derivative $D_\nu$. However we must care
the good behavior of $\Omega_{gh\nu}$ in the manifold index $\nu $.
Then, $E_{\rho ,\nu }^{d}$ in Eq.~(\ref{connection0}) has to be
replaced with a covariant derivative; its respective affine
connection must be local Lorentz invariant not to disturb the
behavior (\ref{connection00}). Thus we are left with the Levi-Civita
connection \cite{Brill,Weinberg}, since it depends just on the
(locally Lorentz invariant) metric tensor. By using
Eq.~(\ref{contorsion}), the Levi-Civita covariant derivative of the
tetrad turns out to be
\begin{equation}
E_{\rho ;\nu }^{d}=E_{\rho ,\nu
}^{d}-\genfrac{\{}{\}}{0pt}{0}{\lambda}{\nu\rho}~E_{\lambda
}^{d}=E_{\rho ,\nu }^{d}-(\Gamma _{\nu \rho }^{\lambda }-K_{~\rho
\nu }^{\lambda })~E_{\lambda }^{d}=E_{\rho ,\nu
}^{d}-(e_{f}^{\lambda }~E_{\rho ,\nu }^{f}-K_{~\rho \nu }^{\lambda
})~E_{\lambda }^{d}=K_{~\rho \nu }^{\lambda }~E_{\lambda }^{d}~
\label{LeviCivitaCont}
\end{equation}
Thus, it is \footnote{ $D_{\nu }(\overline{\psi }\psi )=\partial
_{\nu }(\overline{\psi }\psi )$ because $\overline{\psi }\psi $ is a
local Lorentz invariant; then it is $D_{\nu }\overline{\psi
}=\partial _{\nu }\overline{\psi }-\frac{1}{2}~\Omega _{gh\nu
}~\overline{\psi }~S^{gh}$. Besides, $\overline{D_{\nu }\psi }
=\overline{\partial _{\nu }\psi }+\frac{1}{2} ~\Omega _{gh\nu
}~\overline{S^{gh}\psi }=\partial _{\nu }\overline{\psi }+
\frac{1}{2}~\Omega _{gh\nu }~\psi ^{\dag}~(S^{gh})^{\dag}~\gamma
^{0}=D_{\nu } \overline{\psi }$, since it is $\gamma ^{0}\gamma
^{0}=\mathbf{1}$, $\gamma ^{0}\gamma ^{a \dag}\gamma ^{0}=\gamma
^{a}$, and $(S^{gh})^{\dag }=-\gamma^0 S^{gh}\gamma^0$.}
\begin{equation}
\Omega _{gh\nu }=-\eta _{d[g}~e_{h]}^{\rho }~K_{~\rho \nu }^{\lambda
}~E_{\lambda }^{d}=-K_{ghl}~E_{\nu }^{l}~,~~\ \ \ \ ~\ ~\ \ \ D_{\nu
}\psi =\left( \partial _{\nu }-\frac{1}{2}~K_{ghl}~E_{\nu
}^{l}~S^{gh}\right) \psi \label{connection}
\end{equation}
The set of six independent 1-forms $\mathbf{\Omega}_{gh}=-K_{ghl}
\mathbf{E}^l$ is the Levi-Civita spin connection.
$\mathbf{\Omega}_{gh}$ cannot be made zero by choosing a locally
inertial chart since, as a geometric object, it does not depend on
the coordinates. Instead, $\mathbf{\Omega}_{gh}$ is affected by
local Lorentz transformations, because contorsion $K_{ghl} $ is
tensorial only under \textit{global} Lorentz transformations.
However the local transformations are unable to make $K_{ghl}$ zero,
since they contain just $6$ parameters to make $24$ components zero.
While locally inertial charts are useful for viewing geodesics as
straight lines and bosonic field equations as in Special Relativity,
the effects of the spin connection on the Dirac field cannot be
suppressed by changing coordinates or the tetrad \cite{Arminjon}.
The spin connection term in the Dirac Lagrangian implies a
contribution of gravity to the fermion mass \cite{Schrodinger};
constitutes a form of non-minimal coupling.

\bigskip

Leaving aside the issue of compatibility between the covariantized
Dirac Lagrangian and the automatic conservation (\ref{automatic2}),
which would require a locally Lorentz invariant gauge fixing of the
tetrad, let us pass to examine the consequences of the identity
(\ref{TEGRas}) that reflects the (pseudo-) invariance of TEGR under
local Lorentz transformations of the tetrad. For the original
Lagrangian (\ref{LD}) it is
\begin{equation}
\frac{\delta L_{D}}{\delta E_{\mu }^{a}}~\eta ^{a[g}~E_{\mu
}^{h]}=-E~\mathfrak{T}_{~\lambda }^{\mu }~e_{a}^{\lambda }~\eta
^{a[g}~E_{\mu }^{h]}=-E~\mathfrak{T}^{[gh] }\ . \label{DiracSim}
\end{equation}%
Since $\mathfrak{T}^{gh}$ is non-symmetric (even on-shell, when
$L_D$ vanishes); then the result (\ref{DiracSim}) implies that Dirac
Lagrangian (\ref{LD}) is not compatible with TEGR gravity
\cite{Hayashi1972}. Dirac canonical energy-momentum tensor cannot be
a source of TEGR (or GR). The covariant derivative (\ref{cd}) is
unable to modify this feature; it just adds to $\mathfrak{T}^{gh}$
terms that do not contain derivatives of the Dirac field which will
not alter the non-symmetric character of $\mathfrak{T}^{gh}$.

\section{Conclusions}\label{VI}

We have discussed the identities emerging from the gauge symmetry
(\ref{EHgauge}) in teleparallel theories of gravity; they constitute
the automatic conservation of Eq.~(\ref{automatic}) (equivalently,
(\ref{automatic2}) or (\ref{automatic22})). Scalar and spin 1 matter
are compatible sources in teleparallel gravity since their
energy-momentum tensors are (on-shell) conserved. Even the Dirac
field, which couples to the tetrad rather than the metric, becomes
compatible with the automatic conservation if a (locally) gauge
fixed tetrad is adopted. Contrarily, the Dirac field is not
compatible with the identities coming from the symmetry under local
Lorentz transformations of the tetrad that characterizes TEGR (or
GR), and makes the antisymmetric part of its dynamical equations
vanish (see Eq.~(\ref{TEGRas})). Although the Levi-Civita spin
connection (\ref{connection}) was introduced to endow the Dirac
equation with invariance under local Lorentz transformations, this
does not means that the Dirac Lagrangian thus fits the TEGR gauge
symmetry (\ref{VecTransf}). To emulate this symmetry, the matter
action must be invariant under local Lorentz transformations of the
tetrad \textit{alone}, as happens with scalar and spin 1 matter. The
gauge symmetry of the Dirac action requires, instead, the
transformation of \textit{both} the tetrad and the spinor. This
difference between bosonic and fermionic matter is rooted in the
fact that bosonic matter couples to the metric, which is itself
local Lorentz invariant (no derivatives of the metric are needed in
bosonic Lagrangians; the introduction of the Levi-Civita affine
connection is innocuous indeed). Fermionic matter, in turn, couples
to the tetrad. The common replacement $\gamma^c
e_c^\nu=g^{\mu\nu}\gamma_\mu$ is misleading; it pretends a coupling
to the metric but hides the coupling to the tetrad in $\gamma_\mu$.
In sum, the Levi-Civita spin connection $\Omega_{gh\nu}$ was
introduced to make sense of the Dirac theory in a gravitational
context where the tetrad is determined modulo local Lorentz
transformations; however, the problem of considering fermionic
matter as the source of this type of gravity theories is not solved
in this way. Instead of forcing the coupling between TEGR (or GR)
and Dirac theory, we might consider moving towards teleparallel
theories of gravity preserving the six dof associated with Lorentz
transformations of the tetrad \cite{Okolow}, while keeping the
dynamics of the metric close to that of GR. Only the global Lorentz
symmetry would survive in such case; thus no covariant derivative
$D_\nu$ would be necessary. Current teleparallel theories of
modified gravity would not be appropriate for this purpose because
they contain remnant local Lorentz symmetries
\cite{Ferraro2015,Chen}.

\begin{acknowledgments}
The author thanks Alexey Golovnev and Mar\'{\i}a Jos\'{e} Guzm\'{a}n for helpful
discussions. This work was supported by Consejo Nacional de
Investigaciones Cient\'{\i}ficas y T\'{e}cnicas (CONICET) and Universidad de
Buenos Aires.
\end{acknowledgments}


\begin{thebibliography}{99}

\bibliographystyle{plain}
\bibliography{references.bib}

\bibitem{Noether} E. Noether, Invariante Variationsprobleme, Nachr. d. K\"{o}%
nig. Gesellsch. d. Wiss. zu G\"{o}ttingen, Math-phys. Klasse, 235 (1918)
(English translation in arXiv:physics/0503066).

\bibitem{Sundermeyer} K.~Sundermeyer, Symmetries in Fundamental Physics,
(Springer, Cham, Switzerland, 2014).

\bibitem{Ferraro:2016} R.~Ferraro and M.~J.~Guzm\'{a}n, Hamiltonian
formulation of teleparallel gravity, Phys. Rev. D \textbf{94}
(2016), 104045.

\bibitem{Guzman:2020egp} M.~J.~Guzm\'{a}n and Shymaa Khaled Ibraheem,
Classification of primary constraints for new general relativity in
the premetric approach, Int. J. Geom. Methods Mod. Phys.,\textbf{18}
 supp01, 2140003 (2021).

\bibitem{Hayashi} K.~Hayashi and T.~Shirafuji, New general relativity, Phys.
Rev. D \textbf{19}, 3524 (1979).

\bibitem{Cheng:1988zg} W.~H.~Cheng, D.~C.~Chern and J.~M.~Nester, Canonical
Analysis of the One Parameter Teleparallel Theory, Phys. Rev. D
\textbf{38}, 2656 (1988).

\bibitem{Blixt:2018znp} D.~Blixt, M.~Hohmann and C.~Pfeifer, Hamiltonian and
primary constraints of new general relativity, Phys. Rev. D
\textbf{99}, 084025 (2019).

\bibitem{Blixt:2020ekl} D.~Blixt, M.~J.~Guzm\'an, M.~Hohmann and C.~Pfeifer,
Review of the Hamiltonian analysis in teleparallel gravity, Int. J.
Geom. Meth. Mod. Phys. \textbf{18} supp01, 2130005 (2021).

\bibitem{Cho} Y.~M.~Cho, Einstein Lagrangian as the translational Yang-Mills
Lagrangian, Phys. Rev. D \textbf{14}, 2521 (1976).

\bibitem{Aldrovandi:2013wha} R.~Aldrovandi and J.~G.~Pereira, Teleparallel
Gravity: an Introduction (Springer, Berlin/Heidelberg, 2013).

\bibitem{Gravitation} C.~Misner, K.~Thorne, and J.~A.~Wheeler,
Gravitation (Freeman, San Francisco, 1973).

\bibitem{Golovnev:2017dox} A.~Golovnev, T.~Koivisto and M.~Sandstad, On the
covariance of teleparallel gravity theories, Class. Quantum Grav.
\textbf{34}, 145013 (2017).  

\bibitem{Golovnev:2018gk} A.~Golovnev and T.~Koivisto, Cosmological
perturbations in modified teleparallel gravity models, JCAP
\textbf{11} (2018) 012.

\bibitem{Ong} Y.~C.~Ong and J.~M.~Nester, Counting components in the Lagrange
multiplier formulation of teleparallel theories, Eur. Phys. J. C
\textbf{78}, 568 (2018).

\bibitem{Golovnev:2020nln} A.~Golovnev and M.~J.~Guzm\'an, Nontrivial
Minkowski backgrounds in \textit{f(T)} gravity, Phys. Rev. D
\textbf{103}, 044009 (2021). 

\bibitem{Tong} D.~Tong, The Dirac equation, http://www.damtp.cam.ac.uk/user/tong/qft/four.pdf

\bibitem{Gasperini} M.~Gasperini, Theory of Gravitational Interactions,
(Springer, Cham, Switzerland, 2017).

\bibitem{Fock} V.~Fock, Geometrisierung der Diracschen Theorie des Elektrons,
Zeit. f. Phys. \textbf{57}, 261 (1929).

\bibitem{Kay} B.~S.~Kay, Editorial note to: Erwin Schr\"{o}dinger, Dirac
electron in the gravitational field I, Gen. Relativ. Gravit.
\textbf{52}, 3 (2020).

\bibitem{Beltran} J.~Beltr\'{a}n Jim\'{e}nez, L.~Heisenberg and T.~Koivisto,
Class. Quantum Grav. \textbf{37}, 195013 (2020).  

\bibitem{Brill} D.~R.~Brill and J.~A.~Wheeler, Interaction of neutrinos and
gravitational fields, Rev. Modern Phys. \textbf{29}, 465 (1957).
Erratum: Rev. Modern Phys. \textbf{33}, 623 (1961).

\bibitem{Weinberg} S.~Weinberg, Gravitation and Cosmology: Principles and
Applications of the General Theory of Relativity (Wiley, N.Y.,
1972).

\bibitem{Arminjon} M.~Arminjon, Dirac-Type Equations in a Gravitational Field,
with Vector Wave Function, Found. Phys. \textbf{38}, 1020 (2008).

\bibitem{Schrodinger} E.~Schr\"{o}dinger, Diracsches Elektron im Schwerefeld I,
Sitzungsber. Preuss. Akad. Wiss., Phys.-Math. Kl., 105 (1932)
(English version in Gen. Relativ. Gravit. \textbf{52}, 4 (2020)).

\bibitem{Hayashi1972} K.~Hayashi, Some restrictions on the energy-momentum
tensor of a spin 1/2 particle, Lett. Nuovo Cimento \textbf{5}, 529
(1972).

\bibitem{Okolow} A.~Oko{\l}\'{o}w and J.~\'{S}wie\.{z}ewski, Hamiltonian formulation of a simple
theory of the teleparallel geometry, Class. Quantum Grav.
\textbf{29}, 045008 (2012).

\bibitem{Ferraro2015} R.~Ferraro and F.~Fiorini, Remnant group of local
Lorentz transformations in \textit{f(T)} theories, Phys. Rev. D
\textbf{91}, 064019 (2015).

\bibitem{Chen} P.~Chen, K.~Izumi, J.~M.~Nester, and Y.~C.~Ong, Remnant
symmetry, propagation, and evolution in \textit{f(T)} gravity, Phys.
Rev. D \textbf{91}, 064003 (2015).




\end{thebibliography}
\end{document}